\documentclass[conference]{IEEEtran}
\IEEEoverridecommandlockouts
\usepackage{cite}
\usepackage{amsmath,amssymb,amsfonts}
\usepackage{amsthm}
\usepackage{listings}

\usepackage{algpseudocode}
\usepackage[lined,boxed,commentsnumbered,ruled,linesnumbered,longend]{algorithm2e}
\usepackage{graphicx}
\usepackage{textcomp}
\usepackage{xcolor}
\usepackage{subfigure}
\usepackage{hyperref}
\hypersetup{
    colorlinks=true,
    linkcolor=blue,
    filecolor=blue,      
    urlcolor=blue,
    citecolor=blue,
}
\usepackage{hyperref}
\usepackage[ruled,linesnumbered]{algorithm2e}
\usepackage[misc]{ifsym} 

\usepackage[utf8]{inputenc}
\usepackage{array}
\usepackage{caption}
\usepackage{multirow}
\usepackage{ragged2e}
\usepackage{booktabs}
\usepackage{stmaryrd}
\SetKwRepeat{Do}{do}{while}
\def\BibTeX{{\rm B\kern-.05em{\sc i\kern-.025em b}\kern-.08em
    T\kern-.1667em\lower.7ex\hbox{E}\kern-.125emX}}

\DeclareRobustCommand*{\IEEEauthorrefmark}[1]{%
    \raisebox{0pt}[0pt][0pt]{\textsuperscript{\footnotesize\ensuremath{#1}}}}

\begin{document}

\title{LRScheduler: A Layer-aware and Resource-adaptive Container Scheduler in Edge Computing
}

\author{
\IEEEauthorblockN{
Zhiqing Tang\IEEEauthorrefmark{1},
Wentao Peng\IEEEauthorrefmark{2},
Jianxiong Guo\IEEEauthorrefmark{1,2}, 
Jiong Lou\IEEEauthorrefmark{4},
Hanshuai Cui\IEEEauthorrefmark{1},
Tian Wang\IEEEauthorrefmark{1}, 
Yuan Wu\IEEEauthorrefmark{3},
Weijia Jia\IEEEauthorrefmark{1,2}}
\IEEEauthorblockA{\IEEEauthorrefmark{1}Institute of Artificial Intelligence and Future Networks, Beijing Normal University, Zhuhai, China}
\IEEEauthorblockA{\IEEEauthorrefmark{2}Guangdong Key Lab of AI \& Multi-Modal Data Processing, BNU-HKBU United International College, Zhuhai, China}
\IEEEauthorblockA{\IEEEauthorrefmark{3}State Key Lab of Internet of Things for Smart City, University of Macau, SAR Macau, China}
\IEEEauthorblockA{\IEEEauthorrefmark{4}Department of Computer Science and Engineering, Shanghai Jiao Tong University, Shanghai, China}
\IEEEauthorblockA{zhiqingtang@bnu.edu.cn, wentaopeng@uic.edu.cn, jianxiongguo@bnu.edu.cn, lj1994@sjtu.edu.cn,}
\IEEEauthorblockA{hanshuaicui@mail.bnu.edu.cn, tianwang@bnu.edu.cn, yuanwu@um.edu.mo, jiawj@bnu.edu.cn}
\thanks{This work is supported in part by the National Natural Science Foundation of China (NSFC) under Grant 62302048 and Grant 62272050; in part by Science and Technology Development Fund of Macau SAR under Grant 0158/2022/A; in part by Zhuhai Science-Tech Innovation Bureau under Grant No. 2320004002772, and in part by the Interdisciplinary Intelligence SuperComputer Center of Beijing Normal University at Zhuhai. \textit{(Corresponding author: Yuan Wu.)}}
}

\maketitle

\begin{abstract}
Lightweight containers provide an efficient approach for deploying computation-intensive applications in network edge. The layered storage structure of container images can further reduce the deployment cost and container startup time. Existing researches discuss layer sharing scheduling theoretically but with little attention paid to the practical implementation. To fill in this gap, we propose and implement a Layer-aware and Resource-adaptive container Scheduler (LRScheduler) in edge computing. Specifically, we first utilize container image layer information to design and implement a node scoring and container scheduling mechanism. This mechanism can effectively reduce the download cost when deploying containers, which is very important in edge computing with limited bandwidth. Then, we design a dynamically weighted and resource-adaptive mechanism to enhance load balancing in edge clusters, increasing layer sharing scores when resource load is low to use idle resources effectively. Our scheduler is built on the scheduling framework of Kubernetes, enabling full process automation from task information acquisition to container deployment. Testing on a real system has shown that our design can effectively reduce the container deployment cost as compared with the default scheduler.
\end{abstract}


\section{Introduction}

Emerging as a prominent computing paradigm, edge computing enhances resource availability by deploying applications on edge servers closer to users \cite{shi2016edge}. Containers have emerged as the preferred method for deploying services and applications in edge computing, thanks to their lightweight nature and ease of deployment \cite{tang2022layer,ma2018efficient,fu2020fast}. By utilizing containers on edge servers, applications can significantly reduce the response time and enhance the Quality of Service (QoS). Kubernetes has become the leading tool for container cluster orchestration in cloud data centers \cite{carrion2022kubernetes}, managing the entire lifecycle of containers including deployment \cite{tang2022layer}, migration \cite{tang2023multi}, updates \cite{cui2024latency}, and elastic scaling \cite{brooker2023demand}. Kubernetes offers various scheduling strategies, such as \verb|ImageLocality| and \verb|NodeResourcesBalancedAllocation|, to achieve different goals like selecting nodes with pre-existing container images or those with balanced resource usage \cite{rejiba2022custom}. However, few default scheduling strategies consider the limited bandwidth and storage resources in edge computing.

The default Kubernetes scheduling policies are challenging to use directly in edge computing due to limited resources, geographical distribution, and unstable bandwidth \cite{xing2022h, carrion2022kubernetes}. To address this, container management tools like KubeEdge \cite{xiong2018extend}, K3s \cite{k3s}, and Akraino \cite{akraino} extend Kubernetes to the edge by adding features such as robust management and MQTT support \cite{xiong2018extend}. Additionally, tools like Koordinator \cite{koordinator}, Volcano \cite{volcano}, and Katalyst \cite{katalyst} enhance Kubernetes for distributed scenarios by improving QoS support. However, these tools have neglected the issue of limited bandwidth in edge computing, which makes downloading container images time-consuming \cite{fu2020fast}. Container images are stored in layers, and repeated downloads can be reduced by sharing these layers \cite{gu2023lopo}. Existing researches have explored layer sharing and proposed algorithms for container placement \cite{tang2022layer,gu2021layer}, migration \cite{tang2023multi}, and image downloads \cite{gu2023lopo,lou2022efficient} based on layer sharing. Despite this, a systematic implementation of a layer sharing scheduler is still necessary. Implementing this scheduler in edge environments is crucial to reduce deployment cost for many edge clusters managed by Kubernetes.

Implementing the layer-sharing scheduling algorithm in edge clusters is highly significant and challenging. Using the scheduling framework of Kubernetes \cite{scheduling-framework}, we can create various extension points like filter, score, and bind. The filter extension point eliminates nodes that cannot run the container. The score plugin then ranks the remaining nodes. The scheduler calls each scoring extension point for every node. Finally, the bind extension point binds a container to a node. \textit{However, the first challenge remains on how to automatically obtain and score layer information for nodes.} Most existing work lacks systematic implementation, with some basic schedulers requiring prior knowledge of layer information \cite{fu2020fast}. To fill in such gaps, we develop a custom layer-aware scheduler within the Kubernetes scheduling framework that automatically retrieves and updates layer information from the Docker registry, integrating seamlessly with Kubernetes deployments \cite{deployment}. Layer information is periodically retrieved from the registry and cached locally. The scheduler analyzes the required layer information for new container deployment tasks, gathers existing layer information for each edge node, scores the nodes, and then deploys the container accordingly.

However, using only the layer-aware scheduler will make Kubernetes tend to schedule containers on edge nodes with more layers, leading to higher load on these nodes while others remain underutilized. \textit{This leads to the second challenge, i.e., how to make container scheduling decisions while ensuring load balancing among nodes.} Existing researches have considered the resource utilization when scheduling containers \cite{gunasekaran2020fifer}, including the default scheduling policy \verb|NodeResourcesBalancedAllocation| \cite{configuration}. However, none effectively combine layer sharing to further reduce deployment costs while maintaining load balancing. To address this, we integrate the layer-aware scheduler with certain Kubernetes schedulers \cite{configuration}, calculating a new score by weighting both. Since static weights cannot adapt well to varying loads, we design a Layer-aware and Resource-adaptive scheduler (LRScheduler) with dynamic weights. The LRScheduler dynamically adjusts the layer score weight, lowering it during high load to minimize impact and raising it during low load to decrease download costs and shorten container startup time.

\begin{figure}[!t]
    \centering
    \includegraphics[width=1\linewidth]{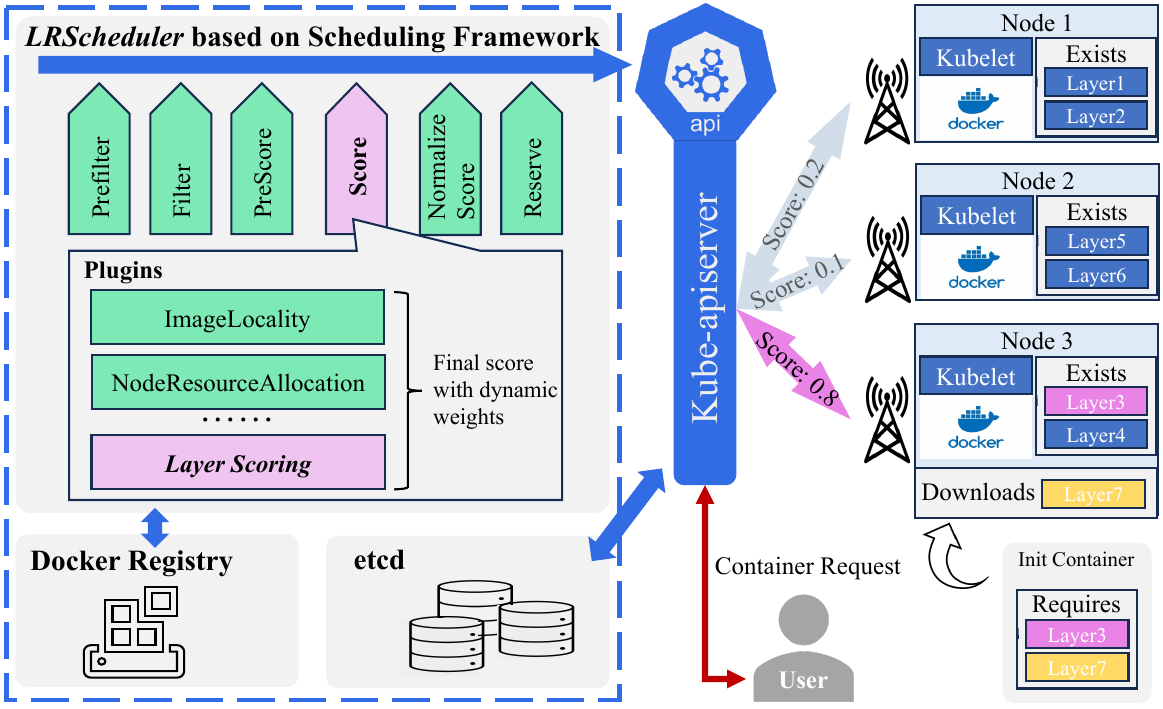} 
    \caption{Overview of LRScheduler}
    \label{fig:system_overview}
\end{figure}

In this paper, we propose and implement a layer-aware and resource-adaptive container scheduler based on the scheduling framework of Kubernetes. As shown in Fig. \ref{fig:system_overview}, LRScheduler implements a custom scoring mechanism based on layer sharing with the help of score extension point, Kubernetes API, etcd, and Kubelet, etc. When a new container request is sent from the user, LRScheduler first retrieves the requested layer information from the user, and the locally stored layer information from each node. Then, it scores nodes using layer information and integrates with default scheduler score to minimize container deployment cost while maintaining load balance. The scheduler dynamically adjusts the weight of layer sharing score and integrates well with various schedulers, providing good scalability. We have implemented this custom scheduler in Kubernetes and verified it in a real cluster environment. Experimental results show that our LRScheduler reduces download cost while considering load balancing.

In summary, the contributions of this paper are as follows:
\begin{enumerate}
    \item We propose and implement a layer-aware scheduler that autonomously calculates the layer sharing score using existing layer data from the node and required layer data from the container. This scheduler can effectively reduce the download cost when deploying containers.
    \item We enhance load balancing by integrating the layer-aware scheduler with the default scheduler and implementing a resource-adaptive weight adjustment algorithm. This approach minimizes the layer download costs during low-load periods while balancing container distribution across nodes under high-load conditions.
    \item We implement our LRScheduler in the Kubernetes system. The experimental results show that our LRScheduler has good scalability. It can effectively reduce the deployment cost of containers and balance the resource load of different nodes.
\end{enumerate}

The rest of the paper is organized as follows. In Section \ref{sec-related-work}, the related work is introduced. System model and problem formulation are described in Section \ref{sec-system-model}. The layer-aware scheduling algorithm is proposed in Section \ref{sec-layer-algorithm}. System implementation details are described in Section \ref{sec-system-implementation} and performance is evaluated in Section \ref{sec-experiments}. Section \ref{sec-conclusion} concludes the paper.

\section{Related Work}
\label{sec-related-work}

\subsection{Layer-aware Container Scheduling}

Layer-aware scheduling research is in its early stages. Rong \textit{et al.} \cite{rong2022exploring} analyze 3735 images from Docker Hub and find that caching image layers on destination servers reduces migration time. Ma \textit{et al.} \cite{ma2018efficient} propose an edge computing platform that uses the layered features of the storage system to reduce the synchronization cost of the file system. Lou \textit{et al.} \cite{lou2022efficient} address the container assignment and layer sequencing problem, proving its NP-hardness, and propose a layer-aware scheduling algorithm. Gu \textit{et al.} \cite{gu2021layer} study a layer aware microservice placement and request scheduling at the edge. Dolati \textit{et al.} \cite{dolati2022layer} address essential aspects of orchestrating services such as downloading and sharing container layers and steering traffic among network functions. Liu \textit{et al.} \cite{liu2022share} study the optimal deployment strategy to balance layer sharing and chain sharing of microservices to minimize image pull delay and communication overhead. 

\subsection{Resource Allocation in Edge Computing}

Significant advancements have been made in resource allocation research for edge computing \cite{wang2021edge,wang2020intelligent}. For example, Xing \textit{et al.} \cite{xing2023harnessing} model the computing resources of the edge nodes uniformly and introduce methods for heterogeneous task classification and recognition. Cai \textit{et al.} \cite{cai2024online} present an explainable online approximation algorithm to optimize resource allocation, balancing model training and inference accuracy. Ouyang \textit{et al.} \cite{ouyang2023dynamic} propose a reactive provisioning approach for hybrid resource provisioning without prior knowledge of future system dynamics. Xu \textit{et al.} \cite{xu2024dynamic} formulate the dynamic parallel multi-server selection and allocation problem to minimize task computing and transmission times. Chen \textit{et al.} \cite{chen2024dynamic} develop an algorithm to minimize system energy consumption while meeting performance requirements for dynamic task offloading and resource allocation. Xu \textit{et al.} \cite{xu2023joint} explore joint channel estimation and resource allocation in Intelligent Reflecting Surface-aided edge computing systems.

However, existing research on layer-aware container scheduling and resource allocation has not effectively combined layer sharing information with load balancing. While some works have made preliminary considerations \cite{tang2022layer, gu2021layer, tang2023multi}, they do not address real system implementation or information retrieval. This paper introduces and implements LRScheduler, an efficient and scalable solution bridging this gap in the current research.

\section{System Model and Problem Formulation}
\label{sec-system-model}

\subsection{System Model}

In edge computing, services are created on specific edge nodes, requiring containers to run. These containers rely on images, which are built from multiple layers.

\textit{Overview}: A set of tasks $\mathbf{K} = \{k_1, k_2, ..., k_{|\mathbf{K}|}\}$ is offloaded from users to edge nodes for processing, where $|\cdot|$ is used to indicate the number of elements in the set, e.g., $|\mathbf{K}|$ is the number of tasks. To handle these tasks, a set of containers $\mathbf{C} = \{c_1, c_2, ..., c_{|\mathbf{C}|}\}$ is deployed on the nodes. Each container requires an image file from the set $\mathbf{M} = \{m_1, m_2, ..., m_{|\mathbf{M}|}\}$. Since requesting a container is equivalent to requesting its corresponding image, and the only difference is a writable container layer, these concepts are unified \cite{zhao2020large, tang2022layer}. Essentially, a task requests a container, which in turn requires specific layers from the set $\mathbf{L} = \{l_1, l_2, ..., l_{|\mathbf{L}|}\}$.

\textit{Edge node}: The set of edge nodes, $\mathbf{N} = \{n_1, n_2, ..., n_{|\mathbf{N}|}\}$, is deployed at the edge of the core network. Each node $n \in \mathbf{N}$ maintains three sets: running containers ${\mathbf{C}}_n(t) \subset \mathbf{C}$, local images ${\mathbf{M}}_n(t) \subset \mathbf{M}$, and local layers ${\mathbf{L}}_n(t) \subset \mathbf{L}$. Additionally, each node has a CPU core number $p_n$, memory capacity $e_n$, bandwidth $b_n$, and storage capacity $d_n$. A node can run a maximum of $C_n$ containers simultaneously.

\textit{Layer}: The set of layers in container $c \in \mathbf{C}$ is ${\mathbf{L}}_c = \left\{x_c^l | l \in \mathbf{L} \right\}$, where $x_c^l = 1$ if container $c$ contains layer $l$, and $x_c^l = 0$ otherwise. The size of layer $l \in \mathbf{L}$ is $d_l$.

\textit{Task}: For each task $k \in \mathbf{K}$ generated by a user at time $t$, the requested CPU resource is $p_k$ and the requested container is $c_k$. After scheduling, the node assigned to this task is $n_k = \left\{u_k^n | n \in \mathbf{N} \right\}$, where $u_k^n = 1$ if task $k$ is scheduled to node $n$, otherwise $u_k^n = 0$.

\subsection{Modeling of Cost and Score}

In edge computing, limited bandwidth and large image sizes result in significant download cost when deploying containers. Compared to this, container startup cost is minimal. Thus, our paper focuses on download cost \cite{tang2022layer}.

For task $k$ requesting container $c$, the requested layers are $\mathbf{L}_c$. At time $t$, the layers stored on edge node $n$ are $\mathbf{L}_n(t)$. The layers from $\mathbf{L}_c$ found on node $n$ are $\mathbf{L}_c \cap \mathbf{L}_n(t)$. The download cost $\mathcal{C}_c^n(t)$ for deploying container $c$ on node $n$ is:
\begin{equation}
\begin{matrix}
    \mathcal{C}_c^n(t) = \sum_{l \in \mathbf{L}_c \setminus \mathbf{L}_n(t)} d_l.
\end{matrix}
\end{equation}
The download time for node $n$ can be obtained as $\mathcal{T}^{k,n} = \frac{\mathcal{C}_c^n(t)}{b_n}$. Moreover, the total size $\mathcal{D}_c^n(t)$ of local layers for node $n$ is:
\begin{equation}
\begin{matrix}
    \mathcal{D}_c^n(t) = \sum_{l \in \mathbf{L}_c \cap \mathbf{L}_n(t)} d_l.
\end{matrix}
\end{equation}
The layer sharing score $\mathcal{S}^{k,n}_{\text{Layer}}(t)$ of node $n$ is obtained as:
\begin{equation}
\begin{matrix}
\label{eq-score-layer}
    \mathcal{S}^{k,n}_{\text{Layer}}(t) = \frac{\mathcal{D}_c^n(t)}{\sum_{l \in \mathbf{L}_c}d_l} \times 100.
\end{matrix}
\end{equation}
Moreover, denote the score of default Kubernetes scheduler as $\mathcal{S}_{\text{K8s}}^{k,n}(t)$, then the weighted score $\mathcal{S}^{k,n}(t)$ can be obtained as:
\begin{equation}
\label{eq-score-final}
    \mathcal{S}^{k,n}(t) = \omega \times \mathcal{S}^{k,n}_{\text{Layer}}(t) + \mathcal{S}_{\text{K8s}}^{k,n}(t),
\end{equation}
where $\omega$ is the weight of layer sharing score. The assigned node $n_k$ for task $k$ is selected with the maximum score:
\begin{equation}
\label{eq-node-selected}
    n_k = \mathop{\arg\max}_{n} \mathcal{S}^{k,n}(t).
\end{equation}

\subsection{Layer-aware Scheduling Problem}

\textit{Constraints}: The constraints are utilized during the Prefilter and Filter plugins during scheduling \cite{scheduling-framework}. The storage capacity of each node should be satisfied:
\begin{equation}
\label{eq-constraints1}
\begin{matrix}
    \mathcal{C}_c^n(t) + \sum_{l \in \mathbf{L}_n(t)}d_l \leq d_n, \quad \forall t, \forall n.
\end{matrix}
\end{equation}
Moreover, the running container number limit is as follows:
\begin{equation}
\label{eq-constraints2}
\begin{matrix}
    |\mathbf{C}_n(t)| \leq C_n.
\end{matrix}
\end{equation}
And each task should only be scheduled to one node:
\begin{equation}
\label{eq-constraints3}
\begin{matrix}
    \sum_{n \in \mathbf{N}} u_k^n = 1, \quad \forall k.
\end{matrix}
\end{equation}

\textit{Problem formulation}: The aim of layer-aware scheduler is to minimize the download cost, i.e., maximize the layer sharing score $\mathcal{S}_{\text{Layer}}^{k,n}(t)$. Therefore, the Layer-Aware Scheduling (LAS) problem can be defined as follows.

\newtheorem*{namedproblemLAS}{Problem LAS}
\newtheorem*{namedproblemLRS}{Problem LRS}

\begin{namedproblemLAS} 
    $\max \mathcal{S}_{\text{Layer}} = \sum_{k \in \mathbf{K}} \mathcal{S}_{\text{Layer}}^{k,n}(t),$
    \label{problem}
    \begin{equation}
    \begin{aligned}
    \label{eq:r}
    \quad \text{s.t.} \quad & \text{Eqs.} \ (\ref{eq-constraints1}), (\ref{eq-constraints2}), (\ref{eq-constraints3}).
    \end{aligned}
    \end{equation}
\end{namedproblemLAS}

By integrating the layer sharing score with other schedulers, the problem can be adapted to different forms, such as when combined with the default scheduler:

\begin{namedproblemLRS} 
    $\max \mathcal{S} = \sum_{k \in \mathbf{K}} \mathcal{S}^{k,n}(t),$
    \label{problem}
    \begin{equation}
    \begin{aligned}
    \label{eq:r}
    \quad \text{s.t.} \quad & \text{Eqs.} \ (\ref{eq-constraints1}), (\ref{eq-constraints2}), (\ref{eq-constraints3}).
    \end{aligned}
    \end{equation}
\end{namedproblemLRS}


\section{Proposed Design of LRScheduler}
\label{sec-layer-algorithm}

\subsection{LRScheduler}

The LRScheduler algorithm, detailed in Algorithm \ref{algorithm_lrs}, takes task $k$ and a set of edge nodes $\mathbf{N}$ as input and outputs the selected node $n_k$ for container deployment.

\begin{algorithm}
	\caption{LRScheduler}\label{algorithm_lrs}
	\SetKwData{Left}{left}\SetKwData{This}{this}\SetKwData{Up}{up}
	\SetKwFunction{Union}{Union}\SetKwFunction{FindCompress}{FindCompress}
	\SetKwInOut{Input}{Input}\SetKwInOut{Output}{Output}
	\Input{$k$, $\mathbf{N}$}
	\Output{$n_k$}
	Initialize $\mathcal{S}^{k,n}(t) \leftarrow 0, \forall k, \forall n$\;
	Update layer information from Registry\;
	\For{$n \leftarrow n_1, n_2, ..., n_{|\mathbf{N}|}$}{
            \tcp{Layer sharing score}
	    Calculate layer sharing score $\mathcal{S}_{\text{Layer}}^{k,n}$ by Eq. (\ref{eq-score-layer})\;
            \tcp{Dynamic weight calculating}
            Calculate weight score $\mathcal{S}_{\text{Weight}}^{k,n}(t)$ by Eq. (\ref{eq-score-weight})\;
            \eIf{$\mathcal{S}_{\text{Weight}}^{k,n}(t) = 1$}{
            $\omega \leftarrow \omega_1$\;
            }{
            $\omega \leftarrow \omega_2$\;
            }
            Get $\mathcal{S}_{\text{K8s}}^{k,n}(t)$ from Kubernetes default scheduler\;
            \tcp{LRScheduler score}
            Calculate the final score $\mathcal{S}^{k,n}(t)$ by Eq. (\ref{eq-score-final})\;
	}
        Select and return the node $n_k$ by Eq. (\ref{eq-node-selected})\;
	\textbf{end}
\end{algorithm}

As shown in Algorithm \ref{algorithm_lrs}, the score is first initialized to zero. Then, for each node $n$, the layer sharing score is calculated by Eq. (\ref{eq-score-layer}). The dynamic weight is calculated as follows. First, the node resource balance score $\mathcal{S}_{\text{STD}}^{k,n}(t)$ is used to measure the resource balance of node $n$ and is defined as:
\begin{equation}
\begin{matrix}
\label{eq-score-resource}
    \mathcal{S}_{\text{STD}}^{k,n}(t) = \frac{|\frac{p_n(t)}{p_n} - \frac{e_n(t)}{e_n}|}{2}.
\end{matrix}
\end{equation}
Then, the score of CPU consumption $\mathcal{S}_{\text{CPU}}^{k,n}(t)$ is obtained as:
\begin{equation}
\begin{matrix}
    \mathcal{S}_{\text{CPU}}^{k,n}(t) = \frac{p_n(t)}{p_n}.
\end{matrix}
\end{equation}
Finally, the dynamic weight score $\mathcal{S}_{\text{Weight}}^{k,n}(t)$ is to decide which weight should be assigned, which is obtained as:
\begin{equation}
\begin{aligned}
\label{eq-score-weight}
    \mathcal{S}_{\text{Weight}}^{k,n}(t) =& \llbracket \mathcal{D}_c^n(t) > h_{\text{size}} \rrbracket \times \llbracket  \mathcal{S}_{\text{CPU}}^{k,n}(t) < h_{\text{CPU}} \rrbracket \\
     & \times \llbracket \mathcal{S}_{\text{STD}}^{k,n}(t) < h_{\text{STD}}\rrbracket,
\end{aligned}
\end{equation}
where $\llbracket \cdots \rrbracket$ is the Iverson bracket, which equals 1 if the condition is met; otherwise, it equals 0.

As shown in lines 8 to 12, if $\mathcal{S}_{\text{Weight}}^{k,n}(t) = 1$, it means the node load is low and the resources are balanced, so we set $\omega \leftarrow \omega_1$, where $\omega_1$ is a relatively large value. Otherwise, $\omega$ is set to a relatively small value $\omega_2$. Then node is selected with the maximum score as shown in line 17.

\subsection{Scalability of LRScheduler}
\label{sec-scalability}

Next, we will discuss the scalability of LRScheduler. As shown in Algorithm \ref{algorithm_lrs}, LRScheduler first scores the layer sharing. Then, it calculates the dynamic weights and gets a final score by weighting it with the scores from other schedulers. By adjusting the method of dynamic weight adjustment, we can conveniently scale LRScheduler to work with any scheduler, minimizing container deployment costs while ensuring the performance of other schedulers. The scalability of LRScheduler mainly reflects in three aspects: the conditions for adjusting dynamic weights, the values of dynamic weights, and the combined schedulers. Their details are as follows.

\textit{Conditions for adjusting dynamic weights.} When calculating $\mathcal{S}_{\text{Weight}}^{k,n}(t)$ in line 7 of Algorithm \ref{algorithm_lrs}, we consider the resource balance, CPU usage, and layer sharing score of the nodes. In fact, other factors can also be taken into account. For example, storage space, memory, even GPU resources, and node availability labels can be further considered to obtain a more refined method for dynamic weight adjustment.

\textit{Values of dynamic weights.} We can extend LRScheduler by adjusting the dynamic weights. For example, we can set different values for $\omega_1$ and $\omega_2$. Additionally, we can add more conditions or piecewise functions, like including a case where $\mathcal{S}_{\text{Weight}}^{k,n}(t)=0.5$. Or we can set a function $\omega = f(\mathcal{S}_{\text{Weight}}^{k,n}(t))$ to adjust the weight.

\textit{Combined schedulers.} Moreover, LRScheduler can be integrated with different Kubernetes schedulers. In the experiments discussed in the next section, we have combined LRScheduler with some default plugins, as shown below:
\begin{enumerate}
    \item \verb|ImageLocality| that prefers nodes with the container images already present.
    \item \verb|TaintToleration| that implements taints and tolerations, reducing deployment priority for tainted nodes.
    \item \verb|NodeAffinity| that implements node selectors and affinity, scoring nodes higher that meet more affinity conditions. Preference is given to nodes that satisfy the specified rules.
    \item \verb|PodTopologySpread| that implements container topology spread by selecting the node with the highest score for each topology pair.
    \item \verb|NodeResourcesFit| that verifies if the node has all the resources requested by the container. The default strategy is \verb|LeastAllocated|.
    \item \verb|VolumeBinding| that verifies if the node can bind the requested volumes, prioritizing the smallest volume that meets the required size.
    \item \verb|InterPodAffinity| that implements inter-Pod affinity and anti-affinity similar to \verb|NodeAffinity|.
\end{enumerate}

Notably, the plugins mentioned above can be enabled or disabled individually, and they can also be combined in various ways to achieve different effects. The main extension point of LRScheduler is the score; by integrating the layer-sharing score into the final score, we can consider layer sharing on top of almost any scheduler and reduce container deployment costs. In summary, LRScheduler has excellent scalability.

\section{System Implementation}
\label{sec-system-implementation}

\begin{figure}[!t]
    \centering
    \includegraphics[width=1\linewidth]{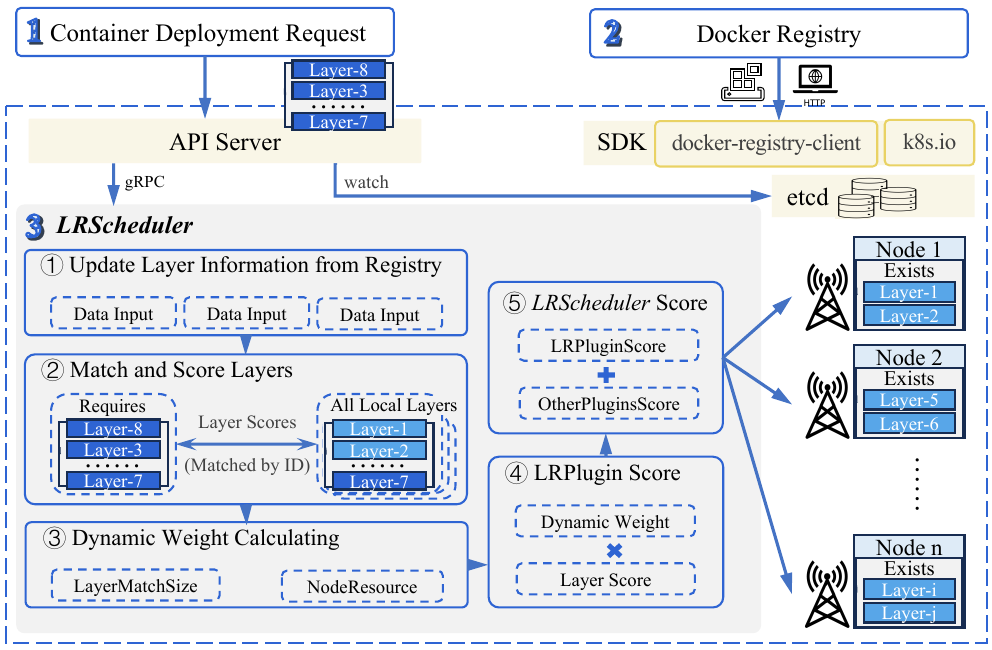} 
    \caption{LRScheduler implementation in Kubernetes system}
    \label{fig:system_implementation}
\end{figure}

As shown in Fig. \ref{fig:system_implementation}, LRScheduler is implemented within the Kubernetes system using the scheduling framework \cite{scheduling-framework}. LRScheduler is deployed to the system using \verb|deployment| \cite{deployment}. First, the user sends a container deployment request, specifying the containers and resource limits, and sets the scheduler to LRScheduler. Upon receiving the request, the Kubernetes API Server invokes LRScheduler for scoring. LRScheduler first updates the layer information from the Registry, then performs layer matching and scoring. After scoring, it calculates dynamic weights and the final score, as detailed in Algorithm \ref{algorithm_lrs}. Once the score is obtained, the Kubernetes API deploys the container to the selected node, completing the entire deployment process. Here are some key details in the implementation process of the LRScheduler as shown in Fig. \ref{fig:system_implementation}.

\subsubsection{Update layer information from Registry} Existing methods cannot automatically retrieve layer information due to challenges in real-time reading and parsing, unstable bandwidth causing connection interruptions in edge computing, and read permission issues from container isolation \cite{fu2020fast}. We address these issues by creating a \verb|goroutine| to periodically fetch all images and their tags from the Docker registry's \lstinline|/v2/_catalog| endpoint. At service start, the \verb|Registry| class initializes. The \lstinline|Registry.Watcher()| method is called and waits for 10 seconds by default to access the registry interface. It filters layer IDs and sizes, stores the data keyed by image name and tag in a JSON file as shown in Listing \ref{listing-data}, and uses this cached file as metadata to compare image sizes through layer information lookup. The retrieved data is formatted into a \lstinline|map[string]ImageMetadata| structure and saved in the \lstinline|cache.json| file.

\begin{lstlisting}[language=Go, caption={Data Structure}, label={listing:data}]
    // Single Layer
    type LayerMetadata struct {
        Size  int64  `json:"size"`
        Layer string `json:"layer"`
    }

    // Single Image
    type ImageMetadata struct {
        Id         string   `json:"id"`
        Name       string   `json:"name"`
        NameWithoutRepo string  `json:"name_without_repo"`
        Tag        string   `json:"tag"`
        TotalSize  int64    `json:"total_size"`
        LayerMetadata   []LayerMetadata  `json:"l_meta"`
    }

    // All Images
    type ImageMetadataLists struct {
       CatchFile string              
       Lists     map[string]ImageMetadata 
    }
\end{lstlisting}

\subsubsection{Match and score layers} Determining the size of the layers and aligning them is challenging. Due to the storage structure, we cannot directly obtain layer size from the image ID. Therefore, we utilize the \verb|cache.json| file as follows:

\begin{enumerate}
    \item The scheduler retrieves scheduled container information from \lstinline|*k8s.io/api/core/v1.Pod|. The image name and tag are accessible via \lstinline|pod.spec.Containers[].Image|.
    \item To obtain the layer sizes from the Registry metadata, we use the image and tag as keys to search the \verb|cache.json| file, returning the layer information \verb|ImageMetadata| for that image.
    \item To calculate the node score, the node information is obtained using the \verb|Handle| method from the base class (\lstinline|framework.ScorePlugin|), specifically \lstinline|k8s.io/kubernetes/pkg/scheduler/framework.Handle|. This includes the node's IP address. By calling the Docker API at \lstinline|http://IP:2375|, all cached images can be retrieved.
    \item Extract layer information from cached image names and tags in the \verb|cache.json| file.
    \item Compare the container layers from step 2 with the cached layers from step 4, extract the matching cached data, and calculate the total cached layer size.
\end{enumerate}

\subsubsection{Dynamic weight calculating} The challenge is how to determine the suitable weights and adjustments based on different needs as discussed in Section \ref{sec-scalability}. The LRScheduler computes dynamic weights through the following steps.

\begin{figure*}[t]
    \centering
    \subfigure[CPU usage]{
    \begin{minipage}[t]{0.275\textwidth}
    \centering
        \includegraphics[width=1\textwidth]{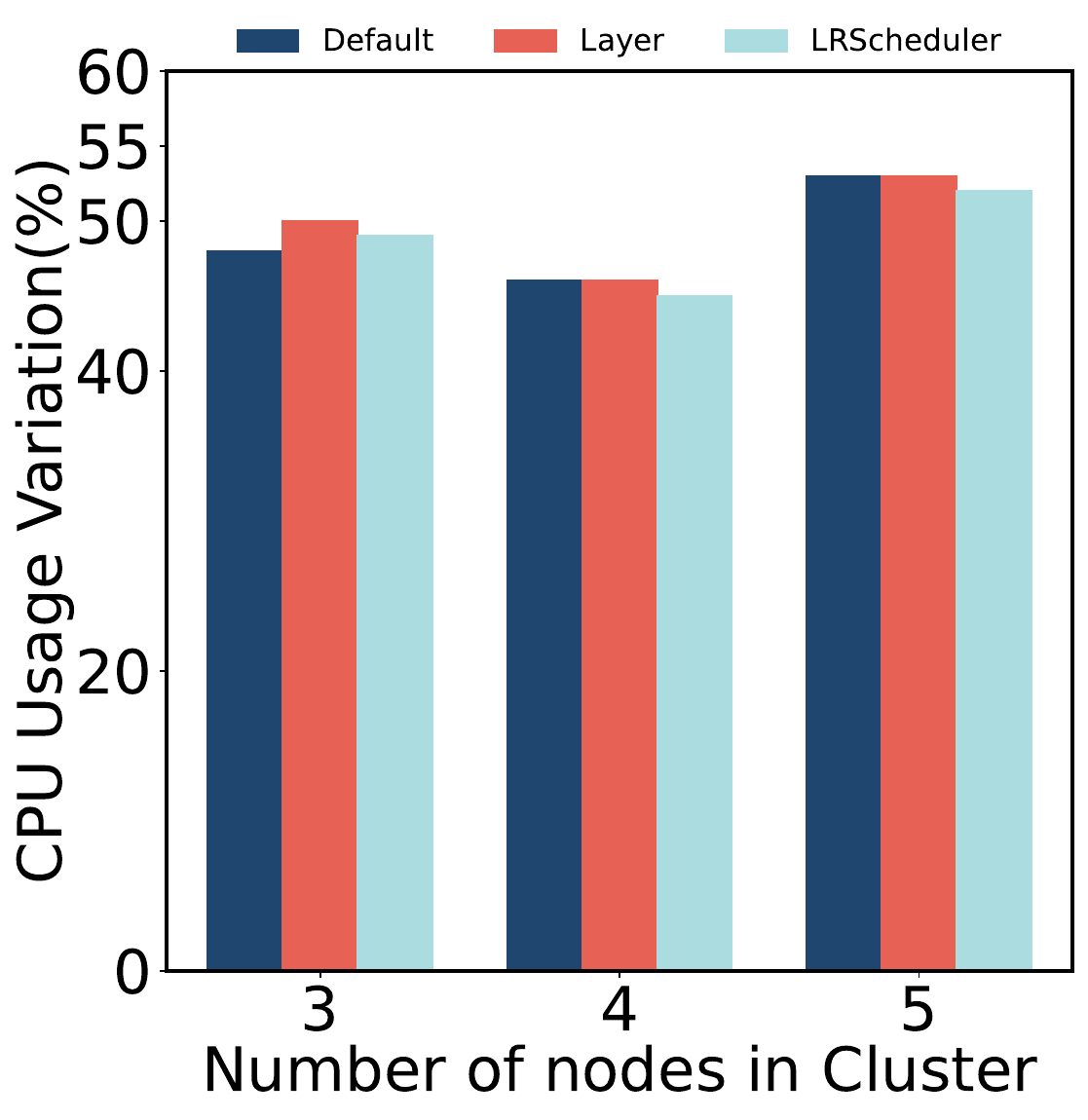}
        \label{fig:CPU-Usage-Variation}
    \end{minipage}
    }
    \hspace{\fill}
    \subfigure[Disk usage]{
    \begin{minipage}[t]{0.275\textwidth}
        \centering
        \includegraphics[width=1\textwidth]{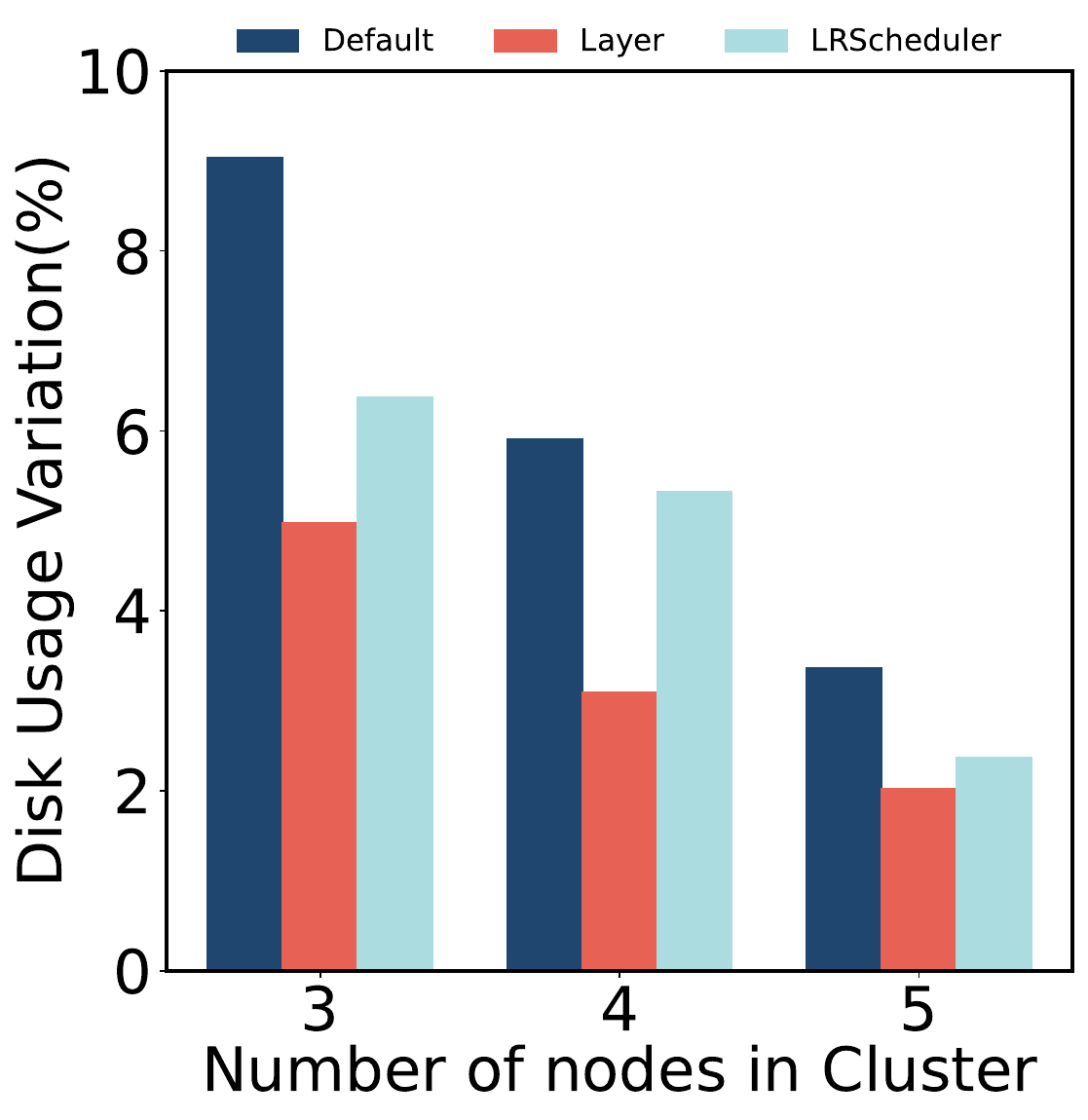}
        \label{fig:Disk-Usage-Variation}
    \end{minipage}
    }
    \hspace{\fill}
    \subfigure[Memory usage]{
    \begin{minipage}[t]{0.275\textwidth}
        \centering
        \includegraphics[width=1\textwidth]{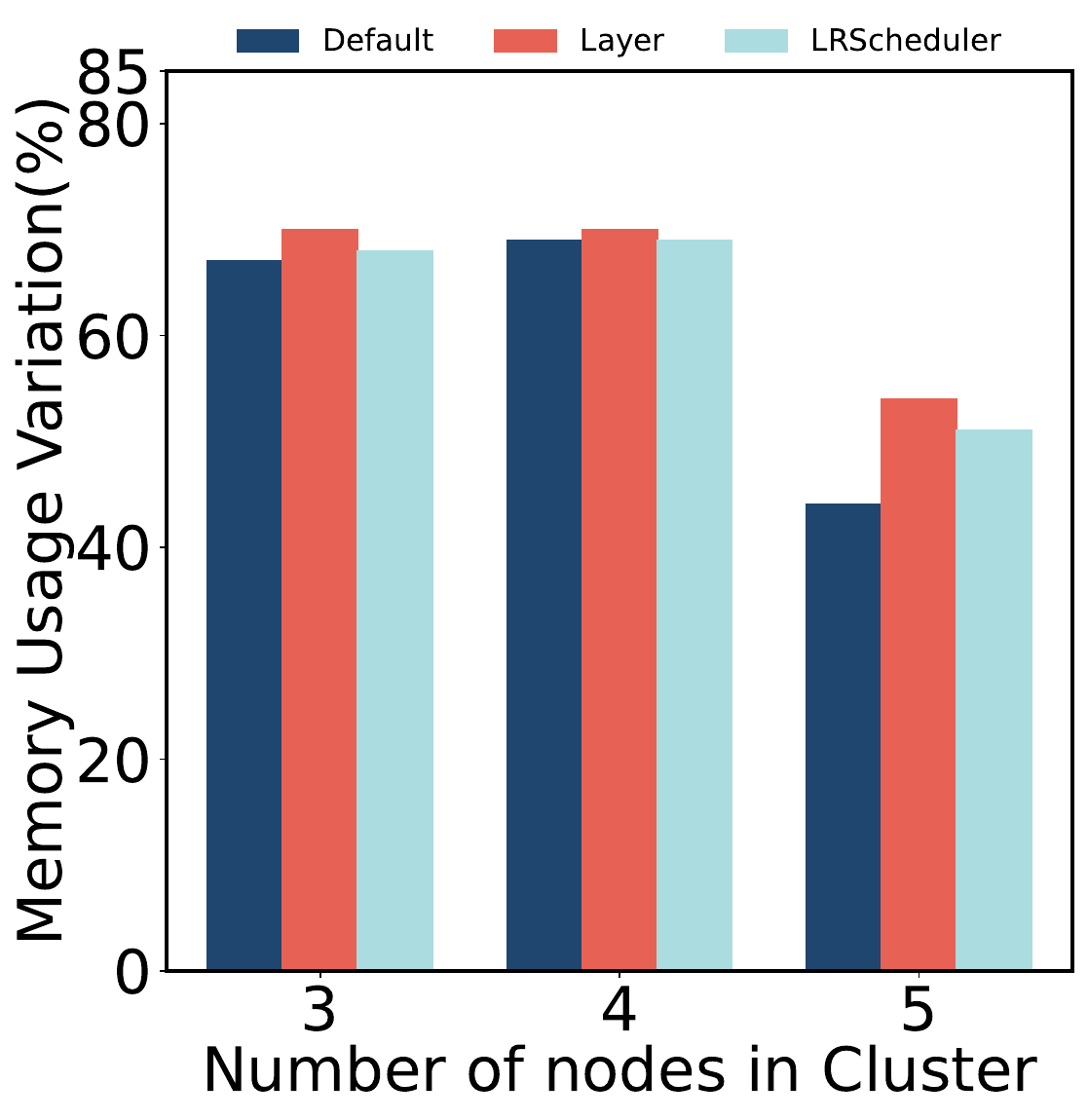}
        \label{fig:Memory-Usage-Variation}
    \end{minipage}
    }
    \hspace{\fill}
    \subfigure[Number of pods]{
    \begin{minipage}[t]{0.275\textwidth}
    \centering
        \includegraphics[width=1\textwidth]{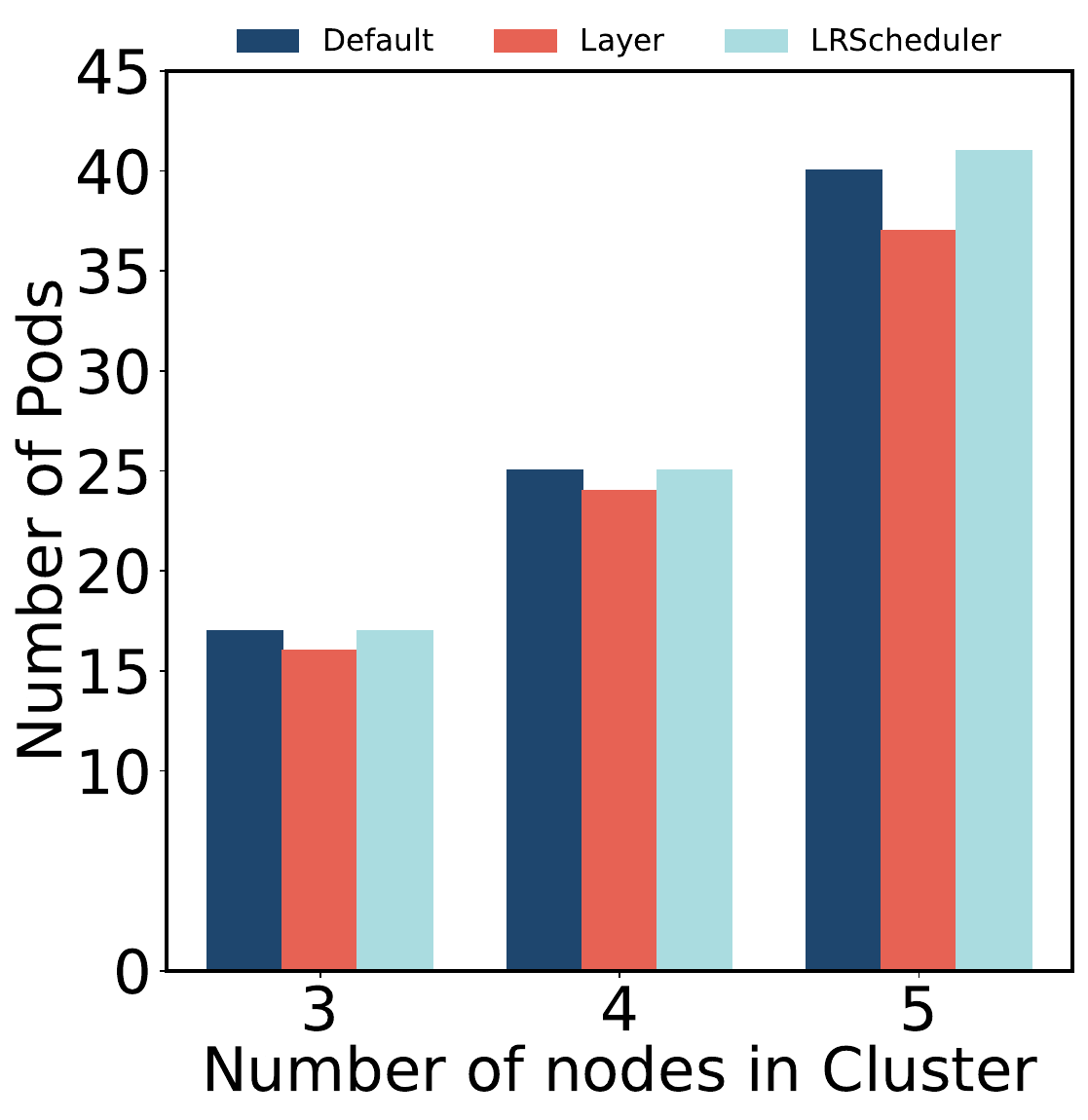}
        \label{fig:MaxNumber-of-Pods}
    \end{minipage}
    }
    \hspace{\fill}
    \subfigure[Download size]{
    \begin{minipage}[t]{0.275\textwidth}
    \centering
        \includegraphics[width=1\textwidth]{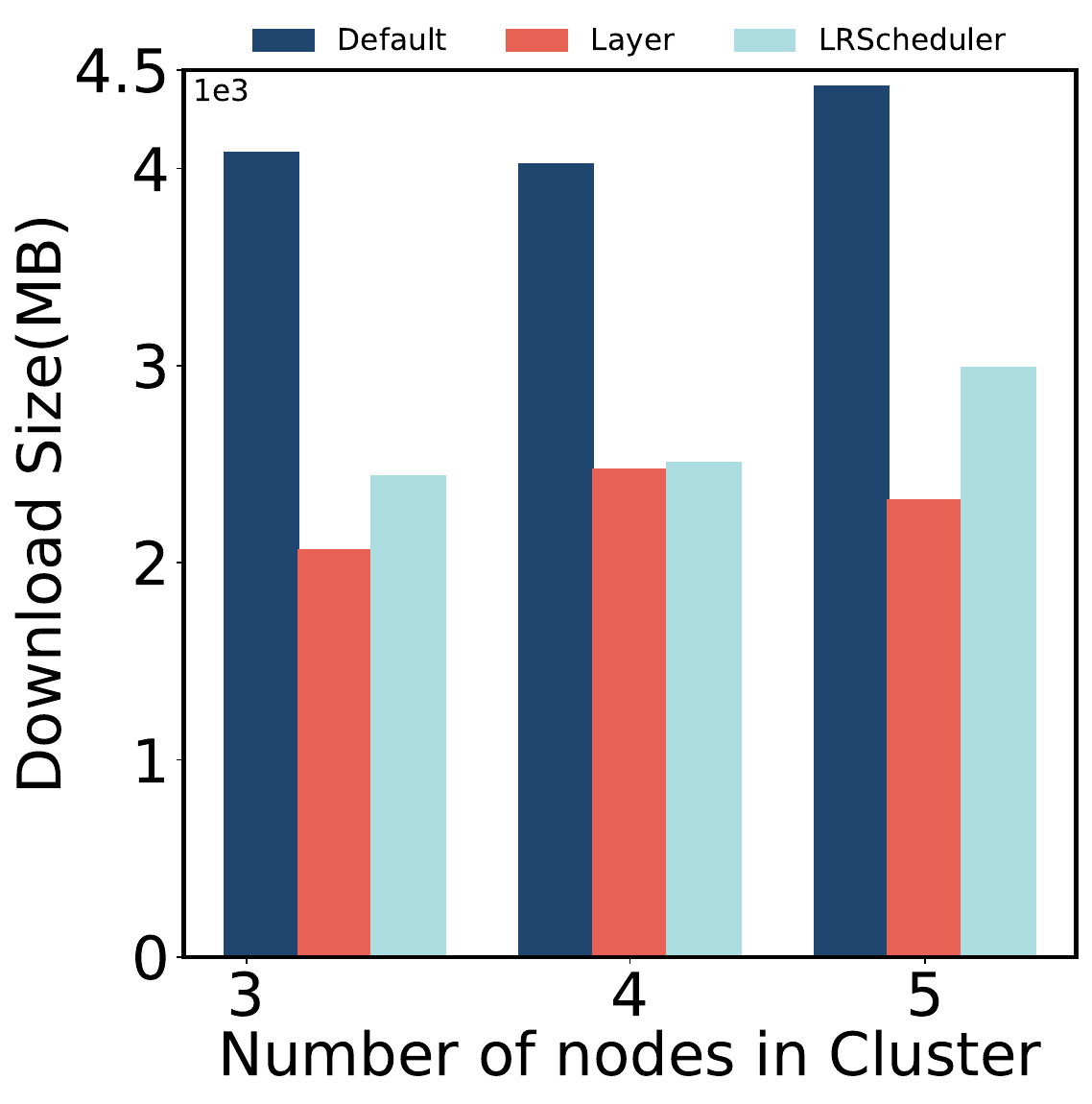}
        \label{fig:Download-Size}
    \end{minipage}
    }
    \hspace{\fill}
    \subfigure[Average standard deviation]{
    \begin{minipage}[t]{0.275\textwidth}
    \centering
        \includegraphics[width=1\textwidth]{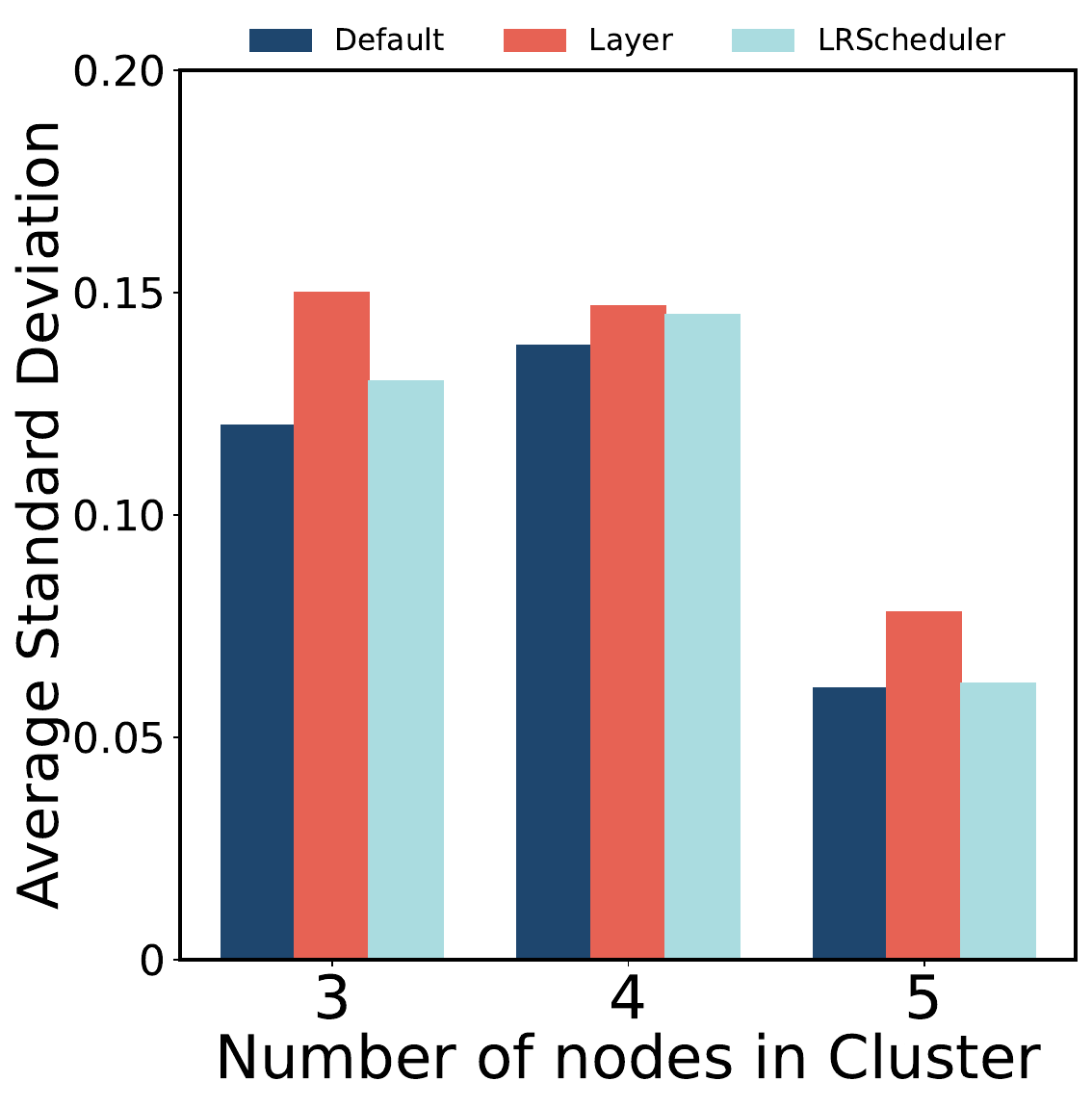}
        \label{fig:Standard-Deviation}
    \end{minipage}
    }
    \caption{Performance with different number of nodes}
    \label{fig-performance-different-nodes}
\end{figure*}

\begin{enumerate}
    \item Using node information (\lstinline|*k8s.io/kubernetes/pkg/scheduler/framework.NodeInfo|), we can access all available resources on the current node and details of all running containers.
    \item Calculate the available CPU and memory percentages by dividing the total requested resources of all containers by the node's available resources. Then, compute the standard deviation (STD).
    \item Return different weights based on Eq. (\ref{eq-score-weight}).
\end{enumerate}

\section{Performance Evaluation}
\label{sec-experiments}

\subsection{Experimental Settings}

To verify LRScheduler, we set up a Kubernetes cluster with 1 master node and 4 worker nodes. All the nodes have Linux CentOS 7 installed. The Kubernetes version used is $v1.23.8$. The container runtime is Docker with version $20.10.8$. The Kubelet and Kube-proxy versions are both $v1.23.8$. All nodes have 4-core CPUs. The master node has 8GB of memory and a 60GB hard drive. Worker node 1 has 4GB of memory and a 30GB hard drive. Worker node 2 has 2GB of memory and a 30GB hard drive. Worker nodes 3 and 4 each have 4GB of memory and a 20GB hard drive. The custom scheduler is implemented in Go language, version $go1.18 linux/amd64$.

We have deployed a private repository using Docker registry. We select some images from Docker Hub and upload them to our private repository, including WordPress, Ghost, GCC, Redis, Tomcat, MySQL, etc. During the experiments, we randomly request these images, setting random CPU and memory limits for each request. Each image consists of several layers, and the information about these layers can be retrieved from the registry. We conduct multiple experiments by deploying different numbers of workers and setting various bandwidth limits. 

We compare LRScheduler with the default scheduler and the Layer scheduler with static weights. The default scheduler enables scheduling plugins as described in Section \ref{sec-scalability}. The Layer scheduler with static weights sets $\omega$ to 4. Additionally, the settings for LRScheduler in the experiment are as follows: $\omega_1 = 2$, $\omega_2 = 0.5$, $h_{\text{size}}=10$, $h_{\text{CPU}}=0.6$, $h_{\text{STD}}=0.16$.

\subsection{Experimental Results}

\textit{Performance with different number of nodes.} We conduct experiments with 3, 4, and 5 edge nodes, as shown in Fig. \ref{fig-performance-different-nodes}. In deployment, Kubernetes operates on Pods, and since our Pods contain only one container, they can be considered equivalent. Fig. \ref{fig:CPU-Usage-Variation} shows that CPU usage across different schedulers and node counts is similar, indicating our scheduler doesn't add extra overhead. Fig. \ref{fig:Disk-Usage-Variation} illustrates disk usage, showing that the Layer scheduler reduces usage by 44\% on average compared to the default scheduler, while LRScheduler achieves a 23\% reduction. This is because LRScheduler optimizes both layer sharing and resource balancing. Fig. \ref{fig:Memory-Usage-Variation} shows minimal differences in memory usage among the three schedulers.

\begin{table*}[t]
\centering
\caption{Performance analysis for 20 containers}
\label{table-performance-analysis}
\begin{tabular}{ccccc|ccccc} \toprule
Container           & Scheduler   & Download Size (MB) & Time (s) & STD   & Container           & Scheduler   & Download Size (MB) & Time (s) & STD    \\ \hline
\multirow{3}{*}{1}  & Default     & 364                & 154      & 0.008 & \multirow{3}{*}{11} & Default     & 43                 & 21       & 0.096  \\
                    & Layer       & $\approx$0                  & $\approx$0        & 0.028 &                     & Layer       & 43                 & 20       & 0.106  \\
                    & LRScheduler & $\approx$0                  & $\approx$0        & 0.027 &                     & LRScheduler & 12                 & 6        & 0.102  \\ \hline
\multirow{3}{*}{2}  & Default     & $\approx$0                  & $\approx$0        & 0.021 & \multirow{3}{*}{12} & Default     & 19                 & 7        & 0.098  \\
                    & Layer       & $\approx$0                  & $\approx$0        & 0.033 &                     & Layer       & 3                  & 1        & 0.133  \\
                    & LRScheduler & $\approx$0                  & $\approx$0        & 0.031 &                     & LRScheduler & 3                  & 1        & 0.108  \\ \hline
\multirow{3}{*}{3}  & Default     & 512                & 232      & 0.046 & \multirow{3}{*}{13} & Default     & 19                 & 8        & 0.105  \\
                    & Layer       & 467                & 193      & 0.064 &                     & Layer       & 21                 & 7        & 0.139  \\
                    & LRScheduler & 464                & 200      & 0.063 &                     & LRScheduler & 11                 & 5        & 0.121  \\ \hline
\multirow{3}{*}{4}  & Default     & 484                & 168      & 0.048 & \multirow{3}{*}{14} & Default     & 384                & 162      & 0.106  \\
                    & Layer       & 140                & 60       & 0.070 &                     & Layer       & 355                & 130      & 0.148  \\
                    & LRScheduler & 142                & 57       & 0.069 &                     & LRScheduler & 330                & 142      & 0.142  \\ \hline
\multirow{3}{*}{5}  & Default     & 116                & 48       & 0.039 & \multirow{3}{*}{15} & Default     & 37                 & 18       & 0.115  \\
                    & Layer       & 110                & 48       & 0.070 &                     & Layer       & 28                 & 12       & 0.155  \\
                    & LRScheduler & 113                & 42       & 0.059 &                     & LRScheduler & 29                 & 10       & 0.147  \\ \hline
\multirow{3}{*}{6}  & Default     & 546                & 228      & 0.045 & \multirow{3}{*}{16} & Default     & 141                & 51       & 0.116  \\
                    & Layer       & 496                & 213      & 0.071 &                     & Layer       & 6                  & 2        & 0.163  \\
                    & LRScheduler & 506                & 219      & 0.063 &                     & LRScheduler & 30                 & 10       & 0.151  \\ \hline
\multirow{3}{*}{7}  & Default     & 123                & 47       & 0.049 & \multirow{3}{*}{17} & Default     & 518                & 189      & 0.124  \\
                    & Layer       & 129                & 43       & 0.076 &                     & Layer       & 99                 & 43       & 0.172  \\
                    & LRScheduler & 130                & 49       & 0.071 &                     & LRScheduler & 189                & 83       & 0.17   \\ \hline
\multirow{3}{*}{8}  & Default     & 165                & 65       & 0.084 & \multirow{3}{*}{18} & Default     & 238                & 91       & 0.151  \\
                    & Layer       & 120                & 49       & 0.085 &                     & Layer       & 208                & 79       & 0.173  \\
                    & LRScheduler & 119                & 45       & 0.082 &                     & LRScheduler & 228                & 84       & 0.171  \\ \hline
\multirow{3}{*}{9}  & Default     & 70                 & 23       & 0.089 & \multirow{3}{*}{19} & Default     & 113                & 51       & 0.153  \\
                    & Layer       & 41                 & 28       & 0.090 &                     & Layer       & 28                 & 10       & 0.175  \\
                    & LRScheduler & 41                 & 14       & 0.090 &                     & LRScheduler & 22                 & 8        & 0.172  \\ \hline
\multirow{3}{*}{10} & Default     & 15                 & 6        & 0.090 & \multirow{3}{*}{20} & Default     & 114                & 43       & 0.160  \\
                    & Layer       & 12                 & 4        & 0.100 &                     & Layer       & 169                & 70       & 0.177  \\
                    & LRScheduler & 44                 & 14       & 0.095 &                     & LRScheduler & 89                 & 33       & 0.17   \\ \bottomrule
\end{tabular}
\end{table*}

\begin{figure}[!t]
    \centering
    \includegraphics[width=0.58\linewidth]{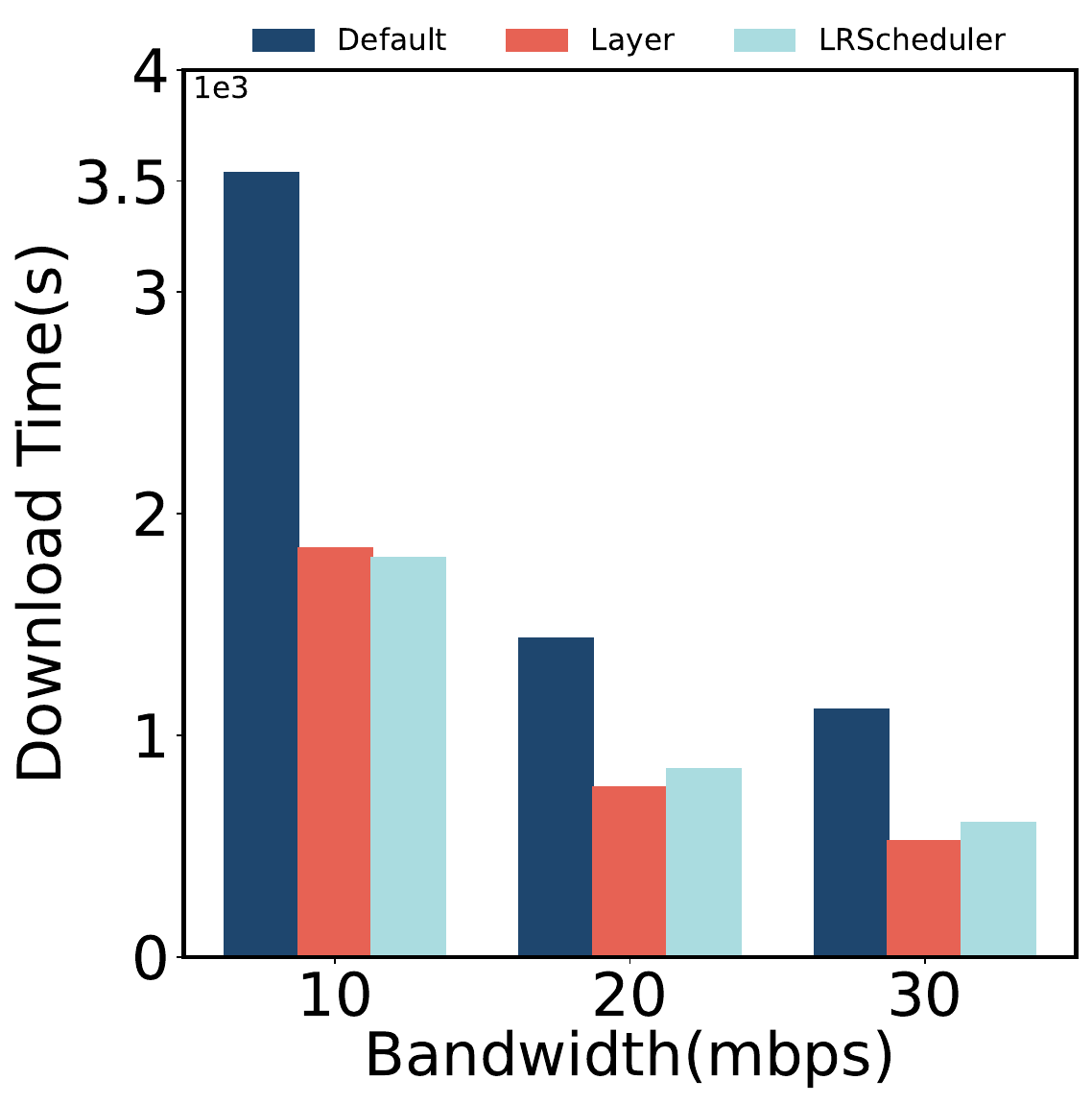} 
    \caption{Performance with different bandwidth}
    \label{fig-download-time-bandwidth}
\end{figure}

\begin{figure}[!t]
    \centering
    \includegraphics[width=0.58\linewidth]{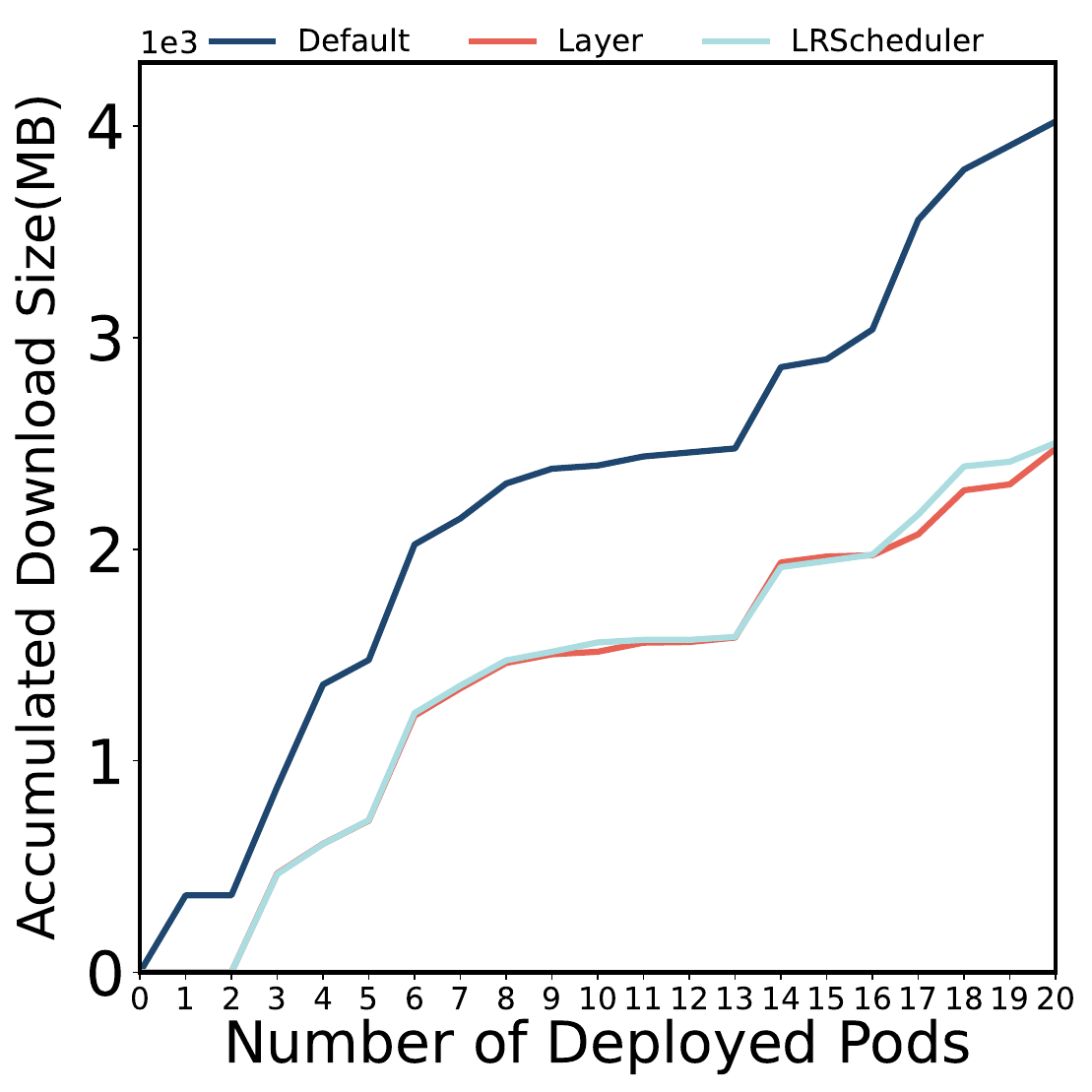} 
    \caption{Accumulated download size for 20 pods}
    \label{fig-accumulated-download}
\end{figure}

Fig. \ref{fig:MaxNumber-of-Pods} illustrates the maximum number of containers that can be deployed on various nodes without image eviction. It shows that LRScheduler can deploy the most containers, highlighting its effective use of layer sharing to minimize redundant disk usage. Fig. \ref{fig:Download-Size} further illustrates that LRScheduler can effectively reduce download cost. Fig. \ref{fig:Standard-Deviation} illustrates LRScheduler can dynamically adjust the weights of layer sharing and score, effectively controlling resource distribution. In summary, LRScheduler effectively reduces download costs while maintaining resource balance. Although the Layer scheduler reduces costs further, it increases the STD (as per Eq. (\ref{eq-score-resource})). Therefore, we can select different schedulers or adjust weights based on our specific needs.

Fig. \ref{fig-download-time-bandwidth} shows the download time at various bandwidths. It is clear that LRScheduler has a more pronounced advantage when the edge network bandwidth is low. Overall, LRScheduler reduces download time by an average of 39\% compared to the default scheduler. Fig. \ref{fig-accumulated-download} illustrates the cumulative download size for deploying 20 different containers. Both Layer Scheduler and LRScheduler demonstrate significantly smaller cumulative download sizes compared to the default scheduler as the number of deployed containers increases. LRScheduler's effectiveness is further demonstrated by its ability to consider additional metrics, such as resource balancing.

Moreover, as shown in TABLE \ref{table-performance-analysis}, we have detailed the download size, download time, and resource balancing (STD) for deploying 20 containers. While LRScheduler may not have the smallest download size at each step, it ultimately results in the lowest total download cost and time, demonstrating its long-term effectiveness despite room for improvement. Besides the scalability discussed in Section \ref{sec-scalability}, reinforcement learning algorithms can also be considered to optimize container deployment costs by accounting for long-term benefits.

\section{Conclusion}
\label{sec-conclusion}

In this paper, we proposed and implemented a layer-aware and resource-adaptive container scheduler for edge computing. First, we designed a scheduler based on layer sharing scores. Second, our scheduler can integrate with the existing Kubernetes scheduler, effectively reducing the download cost during container deployment while meeting other metrics like resource balancing. Finally, we implemented our scheduler, LRScheduler, using Kubernetes' scheduling framework. Through experiments in the Kubernetes system, we validated the effectiveness of LRScheduler. This study demonstrates the feasibility of implementing a layer-sharing-based scheduler in real systems, while highlighting significant opportunities for further optimization. In future work, we will design scheduling algorithms using reinforcement learning and other long-term optimization strategies, and implement them in Kubernetes. Moreover, we will explore cloud-edge collaborative layer sharing to reduce container startup time by transferring layers from other edge nodes.

\bibliographystyle{IEEEtran}
\bibliography{MSN}

\begin{thebibliography}{10}
\providecommand{\url}[1]{#1}
\csname url@samestyle\endcsname
\providecommand{\newblock}{\relax}
\providecommand{\bibinfo}[2]{#2}
\providecommand{\BIBentrySTDinterwordspacing}{\spaceskip=0pt\relax}
\providecommand{\BIBentryALTinterwordstretchfactor}{4}
\providecommand{\BIBentryALTinterwordspacing}{\spaceskip=\fontdimen2\font plus
\BIBentryALTinterwordstretchfactor\fontdimen3\font minus \fontdimen4\font\relax}
\providecommand{\BIBforeignlanguage}[2]{{%
\expandafter\ifx\csname l@#1\endcsname\relax
\typeout{** WARNING: IEEEtran.bst: No hyphenation pattern has been}%
\typeout{** loaded for the language `#1'. Using the pattern for}%
\typeout{** the default language instead.}%
\else
\language=\csname l@#1\endcsname
\fi
#2}}
\providecommand{\BIBdecl}{\relax}
\BIBdecl

\bibitem{shi2016edge}
W.~Shi, J.~Cao, Q.~Zhang, Y.~Li, and L.~Xu, ``Edge computing: Vision and challenges,'' \emph{IEEE internet of things journal}, vol.~3, no.~5, pp. 637--646, 2016.

\bibitem{tang2022layer}
Z.~Tang, J.~Lou, and W.~Jia, ``Layer dependency-aware learning scheduling algorithms for containers in mobile edge computing,'' \emph{IEEE Transactions on Mobile Computing}, vol.~22, no.~6, pp. 3444--3459, 2023.

\bibitem{ma2018efficient}
L.~Ma, S.~Yi, N.~Carter, and Q.~Li, ``Efficient live migration of edge services leveraging container layered storage,'' \emph{IEEE Transactions on Mobile Computing}, vol.~18, no.~9, pp. 2020--2033, 2018.

\bibitem{fu2020fast}
S.~Fu, R.~Mittal, L.~Zhang, and S.~Ratnasamy, ``Fast and efficient container startup at the edge via dependency scheduling,'' in \emph{Proceedings of 3rd USENIX Workshop on Hot Topics in Edge Computing (HotEdge)}, 2020.

\bibitem{carrion2022kubernetes}
C.~Carri{\'o}n, ``Kubernetes scheduling: Taxonomy, ongoing issues and challenges,'' \emph{ACM Computing Surveys}, vol.~55, no.~7, pp. 1--37, 2022.

\bibitem{tang2023multi}
Z.~Tang, F.~Mou, J.~Lou, W.~Jia, Y.~Wu, and W.~Zhao, ``Multi-user layer-aware online container migration in edge-assisted vehicular networks,'' \emph{IEEE/ACM Transactions on Networking}, vol.~32, no.~2, pp. 1807--1822, 2024.

\bibitem{cui2024latency}
H.~Cui, Z.~Tang, J.~Lou, W.~Jia, and W.~Zhao, ``Latency-aware container scheduling in edge cluster upgrades: A deep reinforcement learning approach,'' \emph{IEEE Transactions on Services Computing}, 2024.

\bibitem{brooker2023demand}
M.~Brooker, M.~Danilov, C.~Greenwood, and P.~Piwonka, ``On-demand container loading in $\{$AWS$\}$ lambda,'' in \emph{Proceedings of 2023 USENIX Annual Technical Conference (USENIX ATC 23)}, 2023, pp. 315--328.

\bibitem{rejiba2022custom}
Z.~Rejiba and J.~Chamanara, ``Custom scheduling in kubernetes: A survey on common problems and solution approaches,'' \emph{ACM Computing Surveys}, vol.~55, no.~7, pp. 1--37, 2022.

\bibitem{xing2022h}
T.~Xing, A.~Barbalace, P.~Olivier, M.~L. Karaoui, W.~Wang, and B.~Ravindran, ``H-container: Enabling heterogeneous-isa container migration in edge computing,'' \emph{ACM Transactions on Computer Systems}, vol.~39, no. 1-4, pp. 1--36, 2022.

\bibitem{xiong2018extend}
Y.~Xiong, Y.~Sun, L.~Xing, and Y.~Huang, ``Extend cloud to edge with kubeedge,'' in \emph{Proceedings of 2018 IEEE/ACM Symposium On Edge Computing (SEC)}.\hskip 1em plus 0.5em minus 0.4em\relax IEEE, 2018, pp. 373--377.

\bibitem{k3s}
\BIBentryALTinterwordspacing
K3s: Lightweight kubernetes. [Online]. Available: \url{https://k3s.io}
\BIBentrySTDinterwordspacing

\bibitem{akraino}
\BIBentryALTinterwordspacing
Akraino. [Online]. Available: \url{https://www.lfedge.org/projects/akraino/}
\BIBentrySTDinterwordspacing

\bibitem{koordinator}
\BIBentryALTinterwordspacing
Qos based scheduling system koordinator. [Online]. Available: \url{https://koordinator.sh}
\BIBentrySTDinterwordspacing

\bibitem{volcano}
\BIBentryALTinterwordspacing
Volcano. [Online]. Available: \url{https://volcano.sh}
\BIBentrySTDinterwordspacing

\bibitem{katalyst}
\BIBentryALTinterwordspacing
Katalyst. [Online]. Available: \url{https://gokatalyst.io}
\BIBentrySTDinterwordspacing

\bibitem{gu2023lopo}
L.~Gu, J.~Huang, S.~Huang, D.~Zeng, B.~Li, and H.~Jin, ``Lopo: An out-of-order layer pulling orchestration strategy for fast microservice startup,'' in \emph{Proceedings of 2023 IEEE Conference on Computer Communications (INFOCOM)}.\hskip 1em plus 0.5em minus 0.4em\relax IEEE, 2023, pp. 1--9.

\bibitem{gu2021layer}
L.~Gu, D.~Zeng, J.~Hu, B.~Li, and H.~Jin, ``Layer aware microservice placement and request scheduling at the edge,'' in \emph{Proceedings of 2021 IEEE Conference on Computer Communications (INFOCOM)}.\hskip 1em plus 0.5em minus 0.4em\relax IEEE, 2021, pp. 1--9.

\bibitem{lou2022efficient}
J.~Lou, H.~Luo, Z.~Tang, W.~Jia, and W.~Zhao, ``Efficient container assignment and layer sequencing in edge computing,'' \emph{IEEE Transactions on Services Computing}, vol.~16, no.~2, pp. 1118--1131, 2022.

\bibitem{scheduling-framework}
\BIBentryALTinterwordspacing
Scheduling framework. [Online]. Available: \url{https://kubernetes.io/docs/concepts/scheduling-eviction/scheduling-framework/}
\BIBentrySTDinterwordspacing

\bibitem{deployment}
\BIBentryALTinterwordspacing
Deployments. [Online]. Available: \url{https://kubernetes.io/docs/concepts/workloads/controllers/deployment/}
\BIBentrySTDinterwordspacing

\bibitem{gunasekaran2020fifer}
J.~R. Gunasekaran, P.~Thinakaran, N.~C. Nachiappan, M.~T. Kandemir, and C.~R. Das, ``Fifer: Tackling resource underutilization in the serverless era,'' in \emph{Proceedings of the 21st International Middleware Conference (Middleware)}, 2020, pp. 280--295.

\bibitem{configuration}
\BIBentryALTinterwordspacing
Scheduler configuration. [Online]. Available: \url{https://kubernetes.io/docs/reference/scheduling/config/}
\BIBentrySTDinterwordspacing

\bibitem{rong2022exploring}
C.~Rong, J.~H. Wang, J.~Liu, T.~Yu, and J.~Wang, ``Exploring the layered structure of containers for design of video analytics application migration,'' in \emph{2022 IEEE Wireless Communications and Networking Conference (WCNC)}.\hskip 1em plus 0.5em minus 0.4em\relax IEEE, 2022, pp. 842--847.

\bibitem{dolati2022layer}
M.~Dolati, S.~H. Rastegar, A.~Khonsari, and M.~Ghaderi, ``Layer-aware containerized service orchestration in edge networks,'' \emph{IEEE Transactions on Network and Service Management}, vol.~20, no.~2, pp. 1830--1846, 2022.

\bibitem{liu2022share}
Y.~Liu, B.~Yang, Y.~Wu, C.~Chen, and X.~Guan, ``How to share: balancing layer and chain sharing in industrial microservice deployment,'' \emph{IEEE Transactions on Services Computing}, vol.~16, no.~4, pp. 2685--2698, 2022.

\bibitem{wang2021edge}
T.~Wang, Y.~Liu, X.~Zheng, H.-N. Dai, W.~Jia, and M.~Xie, ``Edge-based communication optimization for distributed federated learning,'' \emph{IEEE Transactions on Network Science and Engineering}, vol.~9, no.~4, pp. 2015--2024, 2021.

\bibitem{wang2020intelligent}
T.~Wang, Y.~Liang, Y.~Zhang, X.~Zheng, M.~Arif, J.~Wang, and Q.~Jin, ``An intelligent dynamic offloading from cloud to edge for smart iot systems with big data,'' \emph{IEEE Transactions on Network Science and Engineering}, vol.~7, no.~4, pp. 2598--2607, 2020.

\bibitem{xing2023harnessing}
T.~Xing, H.~Cui, Y.~Chen, Z.~Luo, B.~Guo, Z.~Yu, X.~Guo, and Y.~Ma, ``Harnessing edge computing resources for accelerating industrial tasks,'' in \emph{Proceedings of 2023 19th International Conference on Mobility, Sensing and Networking (MSN)}.\hskip 1em plus 0.5em minus 0.4em\relax IEEE, 2023, pp. 652--659.

\bibitem{cai2024online}
H.~Cai, Z.~Zhou, and Q.~Huang, ``Online resource allocation for edge intelligence with colocated model retraining and inference,'' in \emph{Proceedings of 2024 IEEE Conference on Computer Communications (INFOCOM)}.\hskip 1em plus 0.5em minus 0.4em\relax IEEE, 2024, pp. 1--9.

\bibitem{ouyang2023dynamic}
T.~Ouyang, K.~Zhao, X.~Zhang, Z.~Zhou, and X.~Chen, ``Dynamic edge-centric resource provisioning for online and offline services co-location,'' in \emph{Proceedings of 2023 IEEE Conference on Computer Communications (INFOCOM)}.\hskip 1em plus 0.5em minus 0.4em\relax IEEE, 2023, pp. 1--10.

\bibitem{xu2024dynamic}
C.~Xu, J.~Guo, Y.~Li, H.~Zou, W.~Jia, and T.~Wang, ``Dynamic parallel multi-server selection and allocation in collaborative edge computing,'' \emph{IEEE Transactions on Mobile Computing}, 2024.

\bibitem{chen2024dynamic}
Y.~Chen, J.~Xu, Y.~Wu, J.~Gao, and L.~Zhao, ``Dynamic task offloading and resource allocation for noma-aided mobile edge computing: An energy efficient design,'' \emph{IEEE Transactions on Services Computing}, 2024.

\bibitem{xu2023joint}
W.~Xu, J.~Yu, Y.~Wu, and D.~H.-K. Tsang, ``Joint channel estimation and reinforcement learning-based resource allocation of intelligent reflecting surface-aided multicell mobile edge computing,'' \emph{IEEE Internet of Things Journal}, 2023.

\bibitem{zhao2020large}
N.~Zhao, V.~Tarasov, H.~Albahar, A.~Anwar, L.~Rupprecht, D.~Skourtis, A.~K. Paul, K.~Chen, and A.~R. Butt, ``Large-scale analysis of docker images and performance implications for container storage systems,'' \emph{IEEE Transactions on Parallel and Distributed Systems}, vol.~32, no.~4, pp. 918--930, 2020.

\end{thebibliography}

\end{document}